\documentclass[]{aa}

\usepackage{graphicx,natbib}

\let\tmp\newinsert
\let\newinsert\newbox
\usepackage{floatrow}
\let\newinsert\tmp

\usepackage{subfigure}
\usepackage{amsmath,amssymb}

\usepackage{txfonts}
\usepackage{pifont}

\usepackage{booktabs}

\usepackage{hyperref}
\hypersetup{colorlinks=true,citecolor=blue,linkcolor=blue,urlcolor=black}

\def\corrk1{correlated-$k$}
\def\kcoeffs1{$k$-coefficients}
\def\kcoeff1{$k$-coefficient}
\def\kdist1{$k$-distribution}
\def\gdist1{$g$-distribution}

\def\rjup{R_\mathrm{Jup}}

\def\rsun{ R_\odot}

\def\co2{CO$_2$}
\def\h2o{H$_2$O}
\def\ch4{CH$_4$}
\def\N2{N$_2$}
\def\nh3{NH$_3$}

\def\ex{\textsl{e}}
\def\d{\mathrm{d}}
\def\ii{\mathrm{i}}
\def\jj{\mathrm{j}}

\def\kkk{k}
\def\kmin{\kkk_\mathrm{min}}
\def\kmax{\kkk_\mathrm{max}}
\def\kgrid{\tilde{\kkk}_\jj}
\def\band{\mathrm{b}}
\def\bb{\bar{\mathrm{b}}}
\def\bun{\mathrm{b}_1}
\def\bdeux{\mathrm{b}_2}
\def\Nb{N_\band}
\def\path{u}
\def\Flux{F}
\def\Fin{\Flux_\mathrm{in}}
\def\Fout{\Flux_\mathrm{out}}
\def\Finb{\Fin^{\band}}
\def\Foutb{\Fout^{\band}}

\def\wb{w^{\band}}
\def\wn{\nu}
\def\dwn{\Delta \wn}
\def\dwnb{\Delta \wn ^{\band}}
\def\trans{\mathcal{T}}
\def\transb{\trans^{\band}}
\def\kcoef{k^{\wn}}
\def\kfunc{\hat{k}}
\def\kfuncb{\hat{k}^\band}
\def\g{g}
\def\gpoints{\g_{\ii}}
\def\gfunc{\hat{g}}
\def\gfuncb{\gfunc^\band}
\def\ggrid{\tilde{\g}_\jj}
\def\ggridb{\tilde{\g}^{\band}_\jj}
\def\Ng{N_\g}
\def\Nk{N_\kkk}
\def\Res{R}
\def\Rsamp{\Res_\mathrm{sp}}
\def\Rf{\Res_\mathrm{fin}}

\newcommand{\balign}[1]{
\begin{align}
#1
\end{align}}

\newcommand{\eq}[1]{Eq.\,(\ref{#1})}

\newcommand{\fig}[1]{Fig.\,\ref{#1}}
\newcommand{\figs}[2]{Figs.\,\ref{#1} and \ref{#2}}
\newcommand{\sect}[1]{Sect.\,\ref{#1}}

\newcommand{\tab}[1]{Table\,\ref{#1}}

\newcommand{\itref}[1]{$^{\ref{#1}}$}

\titlerunning{Spectral binning of precomputed correlated-$k$ coefficients
}
\authorrunning{Leconte et al.}

\begin{document}

\title{
Spectral binning of precomputed correlated-$k$ coefficients
\thanks{I wish to dedicated this paper to the memory of Adam P. Showman and Franck Hersant. You both were, each in your own way, an inspiration to me. Godspeed on your journey among the stars. 
}}

\author{J\'er\'emy Leconte\inst{1}}

\institute{
Laboratoire d'astrophysique de Bordeaux, Univ. Bordeaux, CNRS, B18N, all\'ee Geoffroy 
Saint-Hilaire, 33615 Pessac, France
}
\date{}

\offprints{jeremy.leconte@u-bordeaux.fr}

\abstract{
With the major increase in the volume of the spectroscopic line lists needed to perform accurate radiative transfer calculations, disseminating accurate radiative data has become almost as much a challenge as computing it. 
Considering that many planetary science applications are only looking for heating rates or mid-to-low resolution spectra, any approach enabling such computations in an accurate and flexible way at a fraction of the computing and storage costs is highly valuable. For many of these reasons, the \corrk1 approach has become very popular. Its major weakness has been the lack of ways to adapt the spectral grid/resolution of precomputed \kcoeffs1, making it difficult to distribute a generic database suited for many different applications. Currently, most users still need to have access to a line-by-line transfer code with the relevant line lists or high-resolution cross sections to compute \kcoeff1 tables at the desired resolution.
In this work, we demonstrate that precomputed \kcoeffs1 can be binned to a lower spectral resolution without any additional assumptions, and show how this can be done in practice. We then show that this binning procedure does not introduce any significant loss in accuracy. Along the way, we quantify how such an approach compares very favorably with the sampled cross section approach. This opens up a new avenue to deliver accurate radiative transfer data by providing mid-resolution \kcoeff1 tables to users who can later tailor those tables to their needs on the fly. To help with this final step, we briefly present \texttt{Exo\_k}, an open-access, open-source Python library designed to handle, tailor, and use many different formats of \kcoeff1 and cross-section tables in an easy and computationally efficient way. 
}


\maketitle
\section{Introduction}
Despite the considerable increase in computing power, line-by-line radiative transfer is still considered a computationally intensive task in many cases of interest. One of the reasons for this state of affairs is that recent progress in spectroscopy has multiplied the size of molecular line lists by more than three orders of magnitude \citep{TY12}. We are at a point at which line list themselves, and a fortiori, high-resolution cross sections, represent a data volume that is not trivially handled and exchanged. 

To face this while maintaining flexibility, some have opted for sampled or binned down cross-section tables\footnote{Here, sampled cross sections specifically refer to cross sections that are computed monochromatically but at a resolution that does not necessarily resolve the lines or their shape. Binned-down cross sections are computed on a finely sampled spectral grid and then integrated over a coarser grid, thereby preserving area.} \citep{LTB15,WTR15}. But \citet{GI19} showed that this approach is inaccurate if the resolution used is not sufficiently high, and too computationally expensive if the resolution is high enough (we briefly revisit this question in \sect{sec:sampling} whose results are summarized in \figs{fig:sampling_error}{fig:res_comp}). It seems that for typical resolutions below $\sim$1000, methods designed to group absorptions, such as the so-called \corrk1 method \citep{Liou80,LO91}, are more efficient while remaining accurate \citep{ITD08,ATM17,GI19,ZCK20}. It is especially true for multidimensional models that need a very fast radiative transfer and use only 10-30 bins for the whole spectrum \citep{SFL09,WFS11,LFC13,ATM17}.

For these reasons, it is interesting to directly distribute reference \kcoeffs1 for each molecule instead of high-resolution cross sections. This is currently done for the ExoMol project \citep{CRY20}. This approach decreases the data volume necessary to handle while keeping an optimal accuracy. 
A current drawback with precomputed \kcoeffs1, however, is that the person computing the tables needs to choose a spectral resolution (or wavenumber grid) without knowing exactly what the table will be used for, while \kcoeffs1 are used at their highest potential when tailored to the needed resolution. 

To circumvent this problem we present a way to accurately bin down precomputed \kcoeffs1 to an arbitrary spectral grid without any significant additional loss in accuracy. This method is faster than computing the \kcoeffs1 directly from line-by-line data. It therefore opens up a new way of disseminating accurate, reference molecular opacities: Spectroscopists only need to compute \kcoeffs1 once with a resolution that is higher than the highest resolution they envision their users need (e.g., the resolution of the instrument you want to compare your model to). Users can then easily bin down the data to their spectral grid of choice, possibly even on the fly. 

To help with the latter, we developed \verb"Exo_k"\footnote{See the online documentation for tutorials, example scripts and notebooks, and an extensive and up-to-date description of the (evolving) features of the library: \url{http://perso.astrophy.u-bordeaux.fr/~jleconte/exo_k-doc/index.html} \label{exok_foot}}, an open-source Python 3 library that can directly load precomputed \kcoeff1 tables, change their spectral resolution as well as pressure-temperature grids, and save them back on disk. 
The library can handle many different formats used in various codes: for example, TauREx (\citealt{ACW19}; both pickle and HDF5), petitRADTRANS \citep{MWV19}, Nemesis \citep{ITD08}, ARCIS \citep{MOC20}, Exo transmit \citep{KLO17}, LMD Generic global climate model \citep{WFS11,LFC13}. Several of these formats are provided by the latest ExoMol release \citep{CRY20}. Other formats are being implemented and new ones can be added on request. It is also possible to compute \kcoeff1 tables from high-resolution spectra. One can also combine the opacities of several molecules using the random overlap method \citep{LO91} to create a table for a given atmospheric mix, which is very useful for climate models, for example. Finally, the library includes methods to directly interpolate and combine the opacities, and even compute transmission and emission spectra for 1D planetary atmospheres. Leveraging on the \verb"Numba" library, these operations are performed in a very computationally efficient way. So \verb"Exo_k" can be imported directly into any radiative transfer code to handle molecular opacities easily and efficiently. 

In this article, we first briefly compare the relative accuracy and efficiency of the sampled cross section and \corrk1 approaches in \sect{sec:sampling}. After introducing the necessary concepts and notations, \sect{sec:result} demonstrates rigorously why and how to bin down precomputed \kcoeffs1 to an arbitrary resolution. Then, in \sect{sec:validation}, we validate the numerical algorithm presented in this work and implemented in \verb"Exo_k".

\section{Sampled cross sections versus \corrk1 tables}\label{sec:sampling}

To accurately compare the efficiency of various radiative transfer methods when a given resolution and precision are needed, we briefly estimate the accuracy of atmospheric modeling when using the following types of input radiative data: sampled and binned cross section and \kcoeff1 tables.

We compare this for two different atmospheric configurations and observation geometries:
\begin{itemize}
\item [$\bullet$] Transmission spectrum of a hot Jupiter with a 1000\,K isothermal H$_2$/H$_2$O atmosphere in front of a Sun. 
\item [$\bullet$] Emission spectrum of a pure CO$_2$ Mars-like atmosphere.
\end{itemize}
The parameters for these two cases are given in \tab{tab:params}.

\begin{table}[htbp]
\centering
\caption{Standard parameters used in the our two fiducial cases. A hyphen stands for parameters that are not needed for a given case.}
\begin{tabular}{llrr}
\hline \hline
Case& & Hot Jupiter & Mars\\
\hline
Planet radius & $\rjup$ & 1. & -\\
Surface gravity & m s$^{-2}$ & 10. & 3.72\\
Surface temperature & K & 1000. & 200.\\
Stratospheric temperature & K & 1000. & 100.\\
Adiabatic index & & - & 0.22\\
Surface pressure & Pa &  10$^6$ & 640.\\
Volume mixing ratio of H$_2$O & &  10$^{-3}$ & -\\
Stellar radius & $\rsun$ & 1. & -\\
\hline \hline\\ 
\end{tabular}\label{tab:params}
\end{table}

\begin{figure}[htb]  
\includegraphics[scale=0.7,trim = 0cm 0cm 0.cm 0cm, clip]{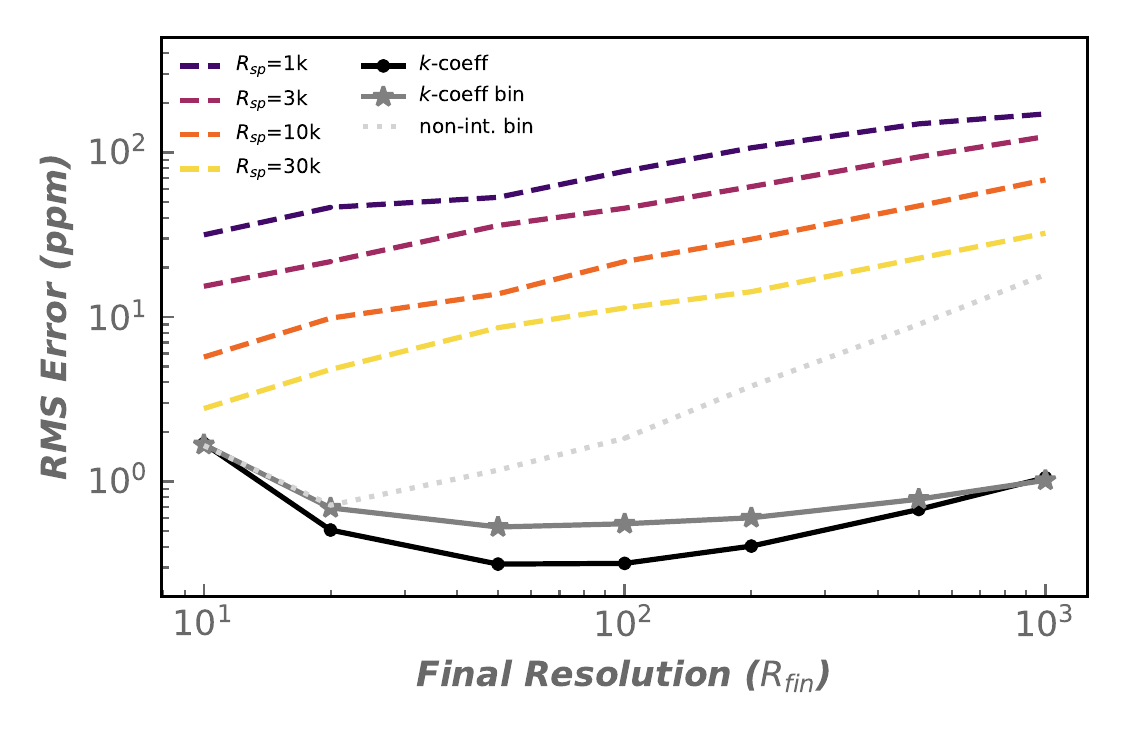}
\caption{Root mean square error on the transit spectrum of our fiducial hot Jupiter over the 1-2\,$\mu$m window as a function of the final observed resolution. Dashed curves represent the error obtained with sampled cross-section tables at various sampling resolution (from top to bottom: $\Rsamp=\{1000,\, 3000,\, 10\,000,\, 30\,000\}$). The spectra are shown in \fig{fig:res_comp}. The black curve with dots represents the error using \kcoeffs1 computed from high-resolution spectra directly at the final resolution. The gray curves show results when using binned down \kcoeffs1 using natural (solid with stars) and non-integer (dotted) binning (See \sect{sec:non-int} for details).}
\label{fig:sampling_error}
\end{figure} 

Each time, we start by computing a reference spectrum using high-resolution monochromatic cross sections (with a typical step of 0.01cm$^{-1}$). For our CO$_2$ atmosphere, we use spectra from \citet{TT17}. For our hot-Jupiter case, we use water spectra produced by the Exomol project with H$_2$ broadening (K. Chubb, private communication) based on the line list produced by \citet{PKZ18}. Collision-induced-absorptions (CIA) are taken from the HITRAN database \citep{RGR11}.

Then, the high-resolution cross sections are sampled at various resolutions ($\Rsamp=\{1\,000,\, 3\,000,\, 10\,000,\, 30\,000\}$) and a new set of spectra are computed at each of these resolutions. 

\begin{figure*}[hp]  
\includegraphics[scale=0.85,trim = 0cm 0cm 0.cm 0cm, clip]{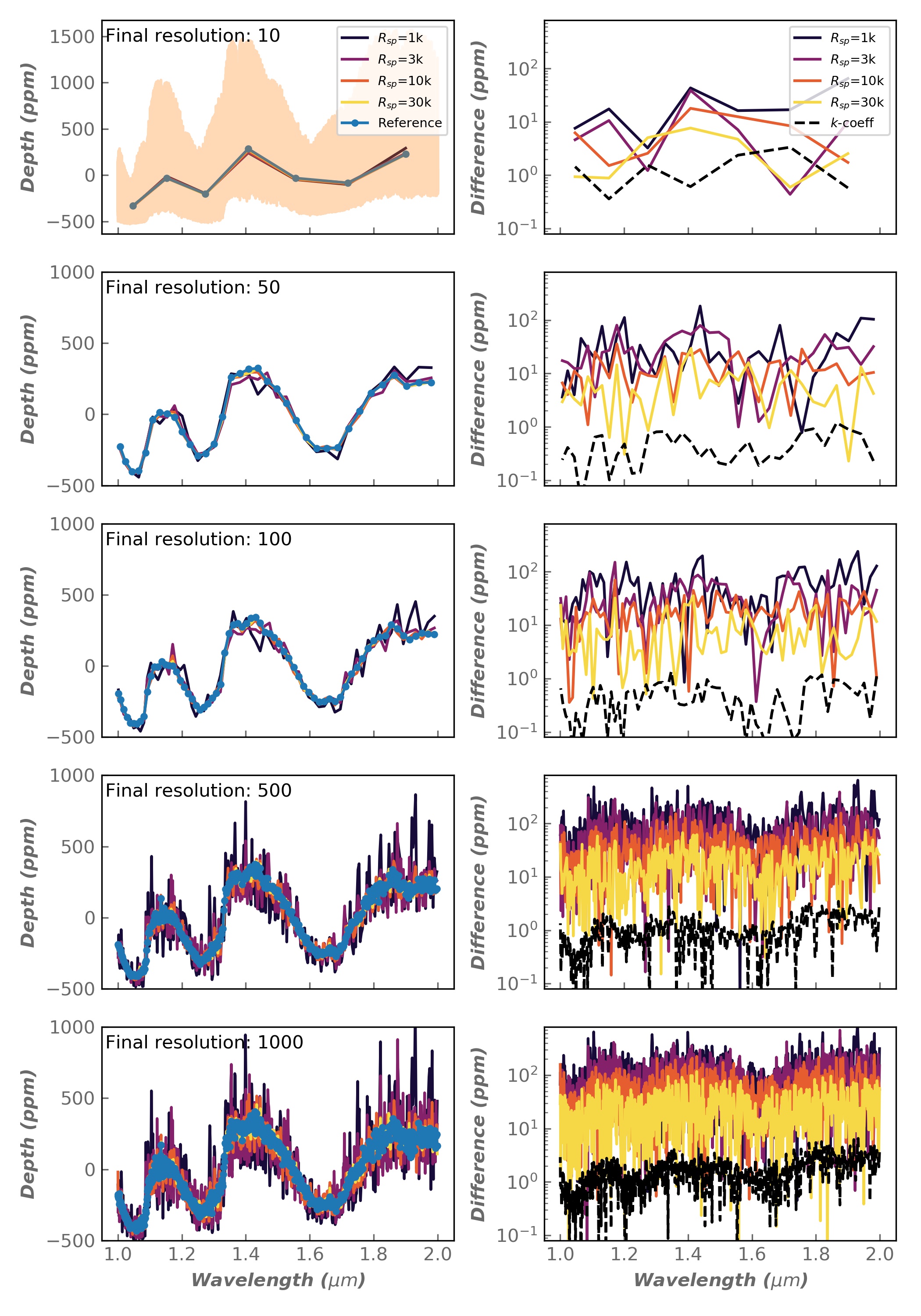}
\caption{Effect of sampling resolution on the transit spectrum of a hot Jupiter. The various rows show the effect for different final observing resolutions ($\Rf$, specified in the top left corner of the panels). The left column shows the transit depth in the 1-2\,$\mu m$ window (with an offset of 11500\,ppm). The various colors correspond to the four sampling resolutions (i.e., the resolution used in radiative transfer calculation before binning; $\Rsamp$) shown in the legend. The curve with dots shows the reference case computed from our very high-resolution cross sections ($\Delta \sigma \approx 0.01$cm$^{-1}$). The high-resolution reference spectrum before binning is shown as the shading in the top left panel. The right column shows the difference (in ppm) between the spectra of the left panel and the reference case. The difference between the calculation with \kcoeffs1 and the reference case is shown with the black dashed curve; the two spectra would be indistinguishable in the left panels. The RMS standard deviation over the spectral region is shown in \fig{fig:sampling_error}.
}
\label{fig:res_comp}
\end{figure*}

\begin{figure*}[htbp]  
\includegraphics[scale=0.85,trim = 0.cm 0cm 0.cm 0cm, clip]{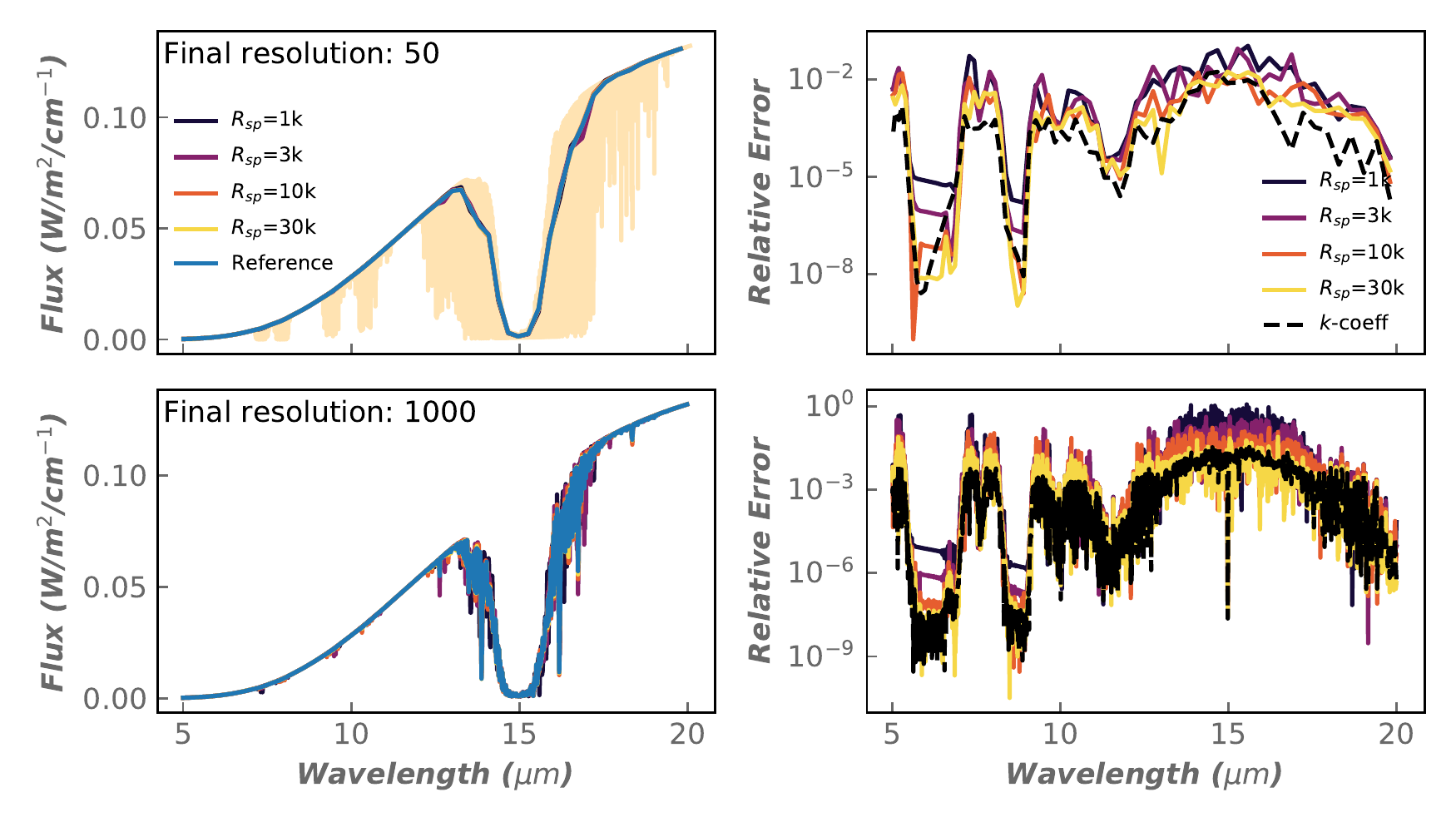}
\caption{Effect of sampling resolution on the emission spectrum of our Mars-like case around the 15$\mu$m CO$_2$ band. The rows show the effect for different final observing resolutions ($\Rf$). The left column shows the flux. The various colors correspond to the four sampling resolutions (i.e., the resolution used in radiative transfer calculation before binning; $\Rsamp$) shown in the legend. The reference case is computed from our very high-resolution cross sections ($\Delta \sigma \approx 0.01$cm$^{-1}$). The high-resolution reference spectrum before binning is shown as the shading in the top left panel. The right column shows the relative error between the spectra of the left panel and the reference case. The error between the calculation with \kcoeffs1 and the reference case is shown with the black dashed curve; the two spectra would be indistinguishable in the left panels.}
\label{fig:res_comp_mars}
\end{figure*}

The spectra obtained from the sampled cross sections are then binned (preserving area) at selected final resolutions ($\Rf=\{10,\, 20,\, 50,\, 100,\, 200,\, 500,\, 1\,000\}$). The transmission spectra computed in the hot-Jupiter case are shown in the left column of \fig{fig:res_comp}. Finally, these spectra are compared to the properly binned reference spectrum (right column of \fig{fig:res_comp}). The root mean squared (RMS) error is shown for each ($\Rsamp$, $\Rf$) in \fig{fig:sampling_error}. The emission spectra computed in the Mars-like case are shown in \fig{fig:res_comp_mars}.

For the \corrk1 approach, \kcoeff1 tables are computed from the high-resolution spectra directly at the final desired resolution (computed with 20 Gauss-Legendre quadrature points in $\g$-space) and compared to the reference spectrum binned at the same resolution. This gives the black dashed curves in \fig{fig:res_comp}. The RMS error is shown by the black curve in \fig{fig:sampling_error}.

For all the computations described above we used a standard algorithm to describe the transmission and emission spectra of a 1D atmosphere that have been implemented in the \verb"Exo_k" library\itref{exok_foot}.

We show that the \corrk1 approach reaches an accuracy that is one to two orders of magnitude better than sampled cross sections, especially for resolutions bigger than 100. The relative drop in accuracy of the \corrk1 method at $\Rf=10$ is due to the low sampling of the CIA absorption, which is treated as constant in each spectral bin. This could be alleviated by applying the random overlap method to the CIA as well. 
To compare numerical efficiency, we focus on the $\Rf$=10 case for which the $\Rsamp=30\,000$ roughly manages to reach an accuracy equivalent to the \corrk1 method. Then, considering that our \kcoeffs1 were computed using 20 quadrature points, the computation is still 150 times faster with the \corrk1 method compared to the sampled cross sections.

\begin{figure*}[htbp]  
\includegraphics[scale=0.85,trim = 0.cm 0cm 0.cm 0cm, clip]{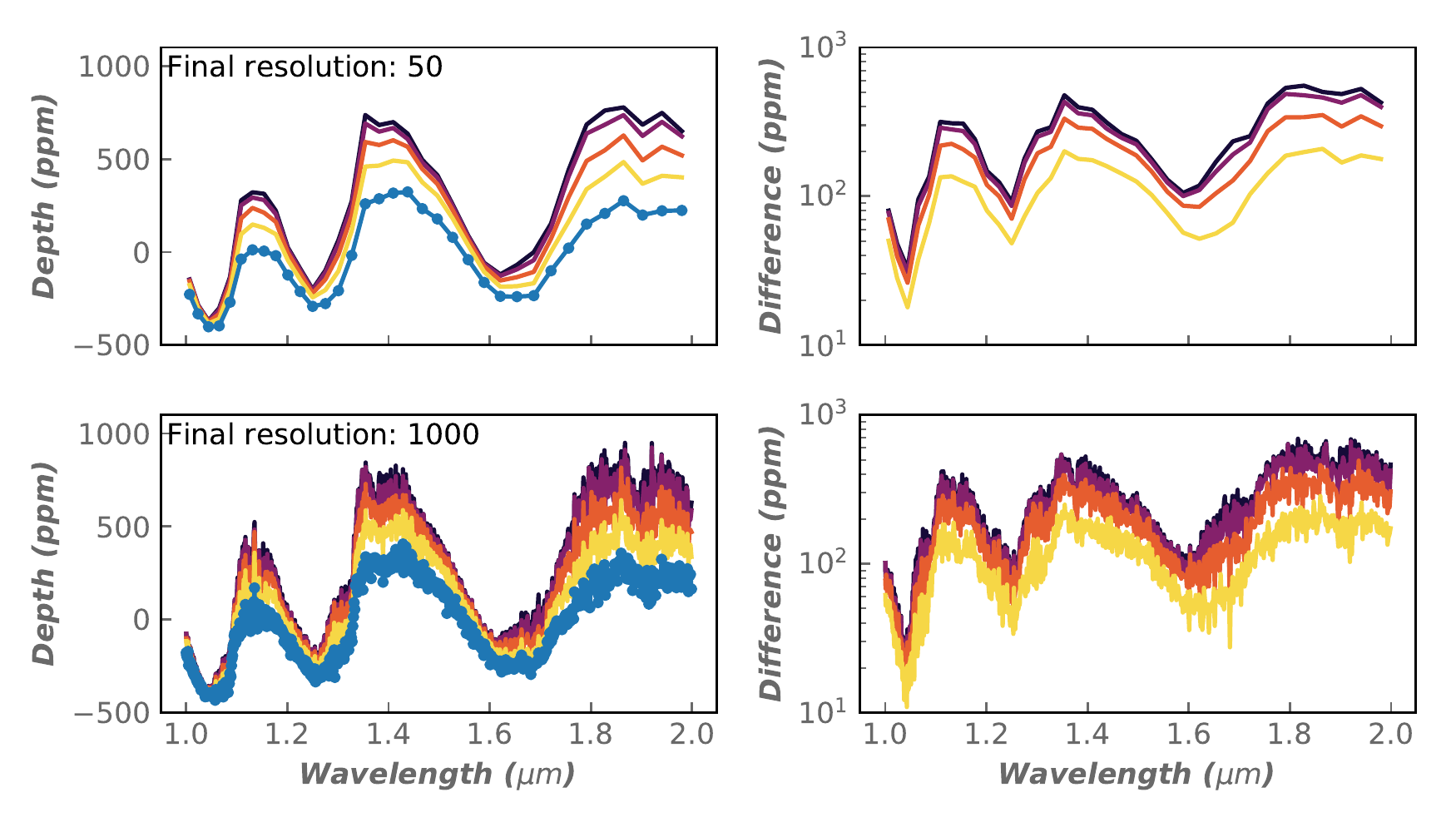}
\caption{Effect of sampling resolution on the transit spectrum of a hot Jupiter using cross sections that have been binned by preserving the total area. The panels can be compared to the results for sampled cross sections in \fig{fig:res_comp} (see the caption for details). The binning systematically overestimates the opacities.
}
\label{fig:res_comp2}
\end{figure*}

We conclude this comparison by mentioning binned-down cross sections -- in which the finely sampled cross sections are integrated over a coarser grid, thereby preserving area. Although it seems the most natural way to bin cross sections, this method performs even worse than the sampled cross sections (by about one order of magnitude; \fig{fig:res_comp2}). This rather counterintuitive result stems from the nonlinear nature of radiative transfer: the opacity dilution due to the binning procedure is not sufficient to increase the transparency of the gas near optically thick lines, while decreasing it in the line wings. Overall, this systematically over estimates the opacity of the atmosphere and creates a systematic bias -- on the order of 100\,ppm in our hot-Jupiter test case.

\section{Method to bin down \kcoeffs1}\label{sec:result}

\subsection{Correlated-$k$ formalism}\label{sec:corrk}

Before demonstrating how to bin down \kcoeffs1, we briefly introduce the basic concepts we need. In the \corrk1 method (see \citealt{Liou80} for details), the wavenumber space, where $\wn$ is the wavenumber, is divided into $\Nb$ spectral bands of width $\dwnb\equiv \wn^\band_{+}-\wn^\band_{-}$. In each band, the transmission through a slab of matter can be written as
\balign{
\transb(\path)=\frac{1}{\dwnb}\int_{\wn^\band_{-}}^{\wn^\band_{+}} \ex^{- \kcoef u} \d \wn= \int_{0}^{1} \ex^{- \kfuncb (\g) u} \d \g \approx  \sum_{\ii=1}^{\Ng} \wb_\ii \ex^{- \kfuncb_\ii u},
}
where $\Ng$ is the number of points (or abscissas) used to discretize the $\g$-space and $\wb$ are the associated set of weights for each band. The symbol $\kfuncb (\g)$ denotes the inverse of the cumulative density function of the opacity within band $\band$ -- the so-called \kdist1 -- and should not be confused with $\kcoef$, the absorption coefficient, which is a function of the wavenumber. The so-called \corrk1 coefficients (or simply \kcoeffs1) are the $\kfuncb_\ii$, which are the discretized version of the \kdist1 ($\kfuncb(\g)$). 

An intuitive way to understand that is to say that now, in band $\band$, the opacity can be described by $\Ng$ representative values ($\kfuncb_\ii$) for each of the $\Ng$ points in $\g$-space ($\gpoints$), each of these values being affected a weight ($\wb_\ii$). Then any radiative transfer calculation can be computed separately in the $\Ng$ bins to be later summed up with the proper weights.

An important, although often forgotten assumption of the \corrk1 formalism is that both the spectral incoming flux impinging on our medium ($\Fin^{\wn}$) and the source function of the latter should be nearly constant within each spectral band \citep{LO91} so that it can be written $\Fin^{\wn}=\Finb/\dwnb$, where $\Finb\equiv \int_{\wn^\band_{-}}^{\wn^\band_{+}} \Fin^{\wn} \d \wn$
is the total flux within the band.
Simply put in the context of a purely absorbing medium, which can be generalized, this ensures that the outgoing flux after a path $u$ is given by
\balign{
\Foutb(\path)\equiv \int_{\wn^\band_{-}}^{\wn^\band_{+}}  \Fin^{\wn}\ex^{- \kcoef (\wn) u} \d \wn&\approx \frac{\Finb}{\dwnb} \int_{\wn^\band_{-}}^{\wn^\band_{+}} \ex^{- \kcoef (\wn) u} \d \wn \nonumber\\
&= \Finb\cdot\transb(\path).
}
This assumption is important in our context because it is the only one that we need to make to be able to combine the \kcoeffs1 of various bands.
Finally, the bolometric flux can be obtained with
\balign{\Flux=\sum_{\band=1}^{\Nb} \Foutb.}

\subsection{Combining \kcoeffs1 of various bands}\label{demo}

To reduce the resolution of precomputed \kcoeffs1, we essentially need to answer the following question:
\begin{itemize}
\item[$\bullet$] Can we determine the \kcoeffs1 $\kfunc^{\bb}_\ii$ of a "super-band" $\bb$ that is the union of two smaller, nonoverlapping "sub-bands"\footnote{We restrict ourselves to the case of the combination of two bands. The generalization to an arbitrary number of bands is trivial.}, $\bun$ and $\bdeux$, for which we know the $k$-coefficients, $\kfunc^{\bun}_\ii$ and $\kfunc^{\bdeux}_\ii$?
\end{itemize}

In this form, the answer is not trivial. However, when computing \kcoeffs1, the crucial mathematical object is not the \kdist1 ($\kfuncb(\g)$) but its inverse, the cumulative density function (CDF) of the opacity within the band (hereafter called the \gdist1, $\gfuncb(\kkk)$) defined as
\balign{\gfuncb(\kkk)\equiv \frac{1}{\dwnb}\int_{\wn^\band_{-}}^{\wn^\band_{+}} H\left(\kkk-\kcoef(\wn)\right) \d \wn, }
where $H(x)$ is equal to 0 where $x$ is positive and 1 elsewhere.

\begin{figure*}[t]  
\includegraphics[scale=0.8,trim = 0cm 0cm 0.cm 0cm, clip]{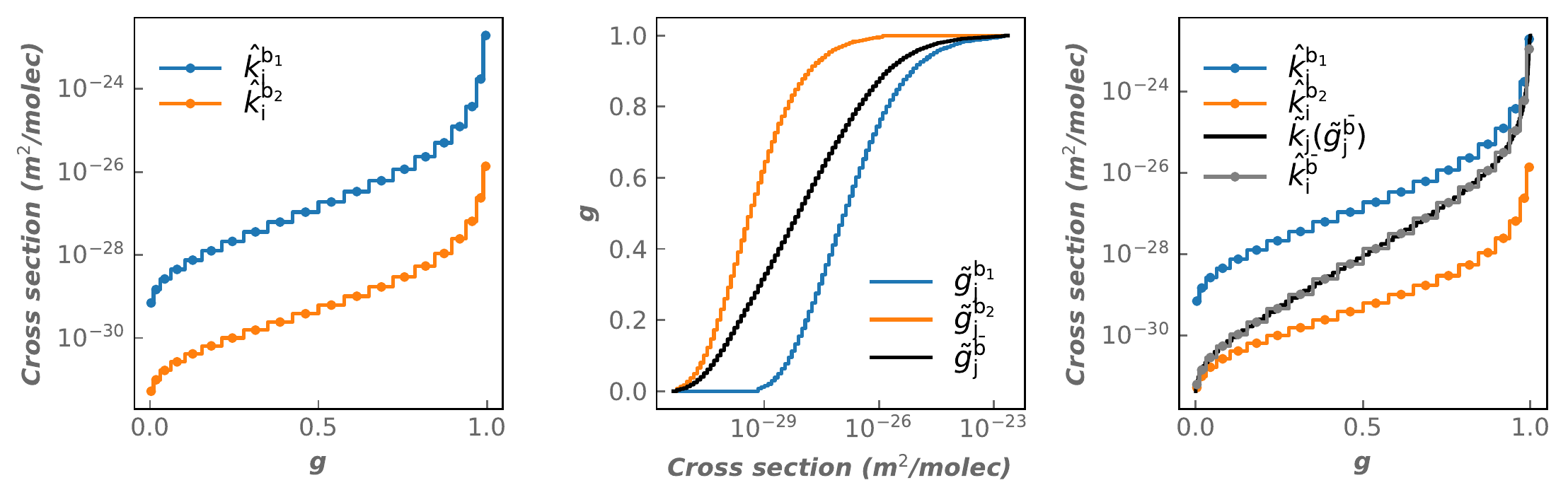}
\caption{Example of the binning process for two bands. Left: \kcoeffs1 on $\Ng=$ 20 Gauss-Legendre quadrature points for two bands ($\kfunc^{\bun}_\ii$ and $\kfunc^{\bdeux}_\ii$). Middle: \gdist1 for the two bands resampled on $\kgrid$, a grid of $\Nk=100$ points between $\kmin$ and $\kmax$ ($\ggrid^{\bun}$ and $\ggrid^{\bdeux}$). The black line indicates the weighted sum of these two $g$-distributions, i.e., the \gdist1 of the super-band ($\ggrid^{\bb}$). Right: The colored lines are shown as in the left panel. The black line indicates the finely sampled \kdist1 ($\kgrid(\ggrid^{\bb})$). The gray line represents the \kdist1 of the super-band re-sampled on the original g-grid with $\Ng=$ 20, i.e., the final \kcoeffs1 ($\kfunc^{\bb}_\ii$).  }
\label{fig:binning}
\end{figure*}

The question can thus be reframed to ask whether we can compute the \gdist1 of the super-band, $\gfunc^{\bb}$, if we know the \gdist1 within the two sub-bands ($\gfunc^{\bun}$ and $\gfunc^{\bdeux}$). This is rather straightforward, because as long as the two smaller bands do not overlap, we have $\dwn^{\bb}=\dwn^{\bun}+\dwn^{\bdeux}$ and 
 \balign{\label{sumCDF}
 \gfunc^{\bb}(\kkk)&\equiv \frac{1}{\dwn^{\bb}}\int_{\wn^{\bb}_{-}}^{\wn^{\bb}_{+}} H\left(\kkk-\kcoef(\wn)\right) \d \wn \nonumber \\
 &= \frac{1}{\dwn^{\bb}}\left(\int_{\wn^{\bun}_{-}}^{\wn^{\bun}_{+}} H\left(\kkk-\kcoef(\wn)\right) \d \wn + \int_{\wn^{\bdeux}_{-}}^{\wn^{\bdeux}_{+}} H\left(\kkk-\kcoef(\wn)\right) \d \wn \right) \nonumber\\
 &=\frac{\dwn^{\bun}}{\dwn^{\bb}} \gfunc^{\bun}(\kkk)+\frac{\dwn^{\bdeux}}{\dwn^{\bb}} \gfunc^{\bdeux}(\kkk). }
Therefore the \gdist1 in the super-band is just the sum of the $g$-distributions in the sub-bands weighted by their spectral extent.

This can be understood more intuitively when discussing in terms of probability. We want to know if $\gfunc^{\bb}(\kkk)$, that is, the probability that a monochromatic ray of light falls in a spectral region where the opacity $\kcoef$ is lower than $\kkk$. To compute this, we have to sum, for each band, the probability of the ray falling in the given sub-band; this probability is $\dwn^{\band}/\dwn^{\bb}$, as long as the incoming spectral flux is constant over the super-band, multiplied by the probability that $\kcoef<\kkk$ within this sub-band, which is $\gfunc^{\band}(\kkk)$. So we recover \eq{sumCDF} as long as we keep the essential assumption of the \corrk1 method that the incoming flux and sources functions do not significantly vary over the wavelength range that we wish to describe by a single distribution.

\subsection{Numerical algorithm}\label{algo}

As shown above, combining the $g$-distributions of various bands in a single \gdist1 for a large band is straightforward. However, to use this in a radiative transfer code, we usually need the \kcoeffs1 ($\kfuncb_\ii$) that are tabulated for fixed $\g$-points ($\gpoints$) with the associated weights ($\wb_\ii$). The difficulty is that although $\gpoints$ can be seen as the values of $\gfuncb$ in band $\band$ for the points located at $\kfuncb_\ii$ in opacity space, the distribution $\gfuncb$ is sampled at different locations in opacity space in each sub-band. 

To circumvent this problem, we recommend the following approach, which is the one implemented in \verb"Exo_k". The whole process is illustrated in \fig{fig:binning} with only two bands, but can be carried out for any number of bands. For the moment, however, we restrict ourselves to the case in which the super-band $\bb$ is composed of an integer number of sub-bands (hereafter, natural binning). Once these sub-bands have been identified, we determine 
\balign{\kmin=\underset{\ii, \,\band \in \bb}{\mathrm{min}}\left\{\kfuncb_\ii\right\} \ \ \mathrm{and}\ \ \kmax=\underset{\ii, \,\band \in \bb}{\mathrm{max}}\left\{\kfuncb_\ii\right\}
}
and compute an evenly log-distributed grid of $\Nk$ points ($\kgrid$) between these two values. Then, for each sub-band, we interpolate $\gpoints(\kfuncb_\ii)$ on this grid, which yields $\Nk$ values per band, $\ggridb$, representing all the $\gfuncb$ on the same grid for all bands. 
It is thus now easy to compute the global \gdist1 which is given by
\balign{\label{weighting}
\ggrid^{\bb}=\sum_{\band\in\bb} \frac{\dwnb}{\dwn^{\bb}} \ggridb.
}
This generally yields an oversampled version of the distribution because we choose $\Nk>\Ng$ (the black curve in the middle panel of \fig{fig:binning}). Recomputing the \kcoeffs1 is thus now simply a matter of resampling $\kgrid(\ggrid^{\bb})$ onto our original $\g$-grid ($\gpoints$) -- going from the black to the gray curve in the right panel of \fig{fig:binning} -- as discussed in \citet{LO91} and \citet{ATM17}. However, unlike those authors, we think that, to be consistent with the usual quadrature rules, the finely sampled \kdist1 should not be weight-averaged over the final $\g$-bins, but sampled at the precise abscissa of the $g$-point determined by the quadrature rule used, following \citet{ITD08}.

As long as we resample on the original grid, we keep the original quadrature with $\Ng$ points, and the weights are left unchanged. However, it is still possible at this point to use different $\g$-points, especially if we need to use less $g$-points for computational reasons (as is often the case with 3D climate models). For this, we can simply resample $\kgrid(\ggrid^{\bb})$ on a different $\g$-grid and use the relevant weights, which is an option of \verb"Exo_k". In general, \verb"Exo_k" works for any choice of $\g$-space sampling specified by the user.


\subsection{Non-integer binning}\label{sec:non-int}

Up to now, we only considered the case in which the final super-band encompasses an integer number of sub-bands. This is required for our demonstration to be exact without any further assumptions. 

In practice, it is still possible to bin \kcoeffs1 on a spectral grid that does not exactly respect the limits of the sub-bands (hereafter, non-integer binning). To do that, we replace the spectral extent, $\dwnb$, of the two bands at the boundaries of the super-band in \eq{weighting} by their spectral extent inside the super-band. This is equivalent to binning a high-resolution spectrum to a low-resolution grid using a weighted average. In theory, this would remain perfectly accurate if the statistical distribution of the absorption coefficient were constant within the sub-band. As this is not true in general, we test the accuracy of this approach in the next section.

\section{Validation}\label{sec:validation}

To validate our approach, we compute spectra in the two cases detailed in \sect{sec:sampling}. As we do not want to test the accuracy of the \corrk1 approach itself, which has already been done before, but whether the binning introduces additional errors, the comparison proceeds as follows:
\begin{itemize}
\item[$\bullet$] Starting from very high-resolution spectra, we compute a first reference set of tables of \kcoeffs1 with resolutions ranging from $\Res=1\,000$ to $\Res=5$ and a $\g$-space sampled using 20 Gauss-Legendre quadrature points. The exact $\Res$ values used do not matter. However, we make sure that the wavenumber points in the low-resolution grids can be found in the high-resolution grid to use only natural binning.
\item[$\bullet$] Starting from the $\Res=1\,000$ table of \kcoeffs1\footnote{This choice is motivated by the resolution chosen by the ExoMol project for their \kcoeff1 tables \citep{CRY20}} we just computed, we use the binning approach described above to compute a new set of ten tables of \kcoeffs1 with the exact same resolutions and wavenumber grids as the reference set. 
\item[$\bullet$] The emitted or transmitted flux of our fiducial atmosphere is computed at each resolution with both the reference and the binned \kcoeffs1 and the results are compared two by two. 
\end{itemize}

\begin{figure}[htbp] 
 \sidecaption\centering
\subfigure{ \includegraphics[scale=0.9,trim = 0cm 1.2cm 0.cm 0.cm, clip]{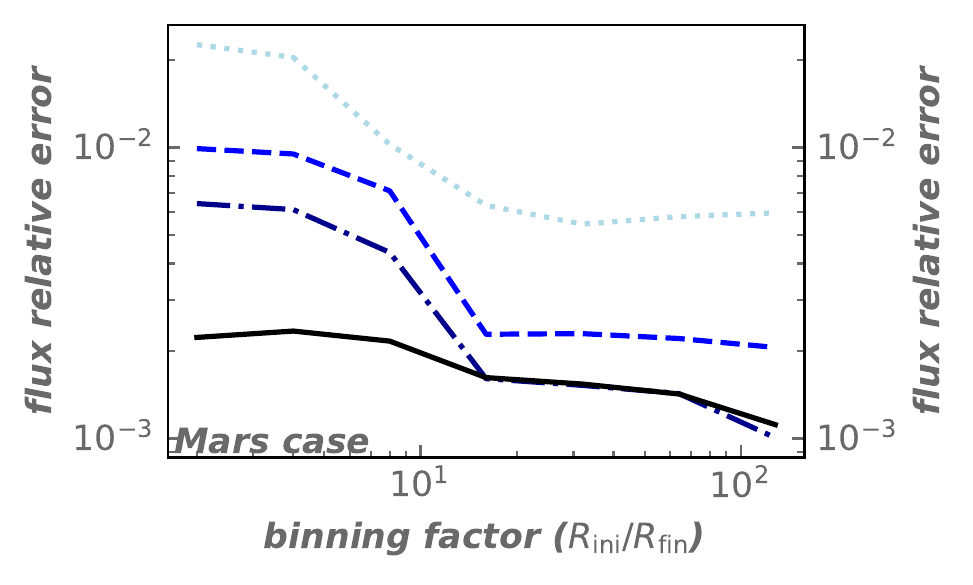} }
\subfigure{ \includegraphics[scale=0.9,trim = .0cm .cm 0.cm .cm, clip]{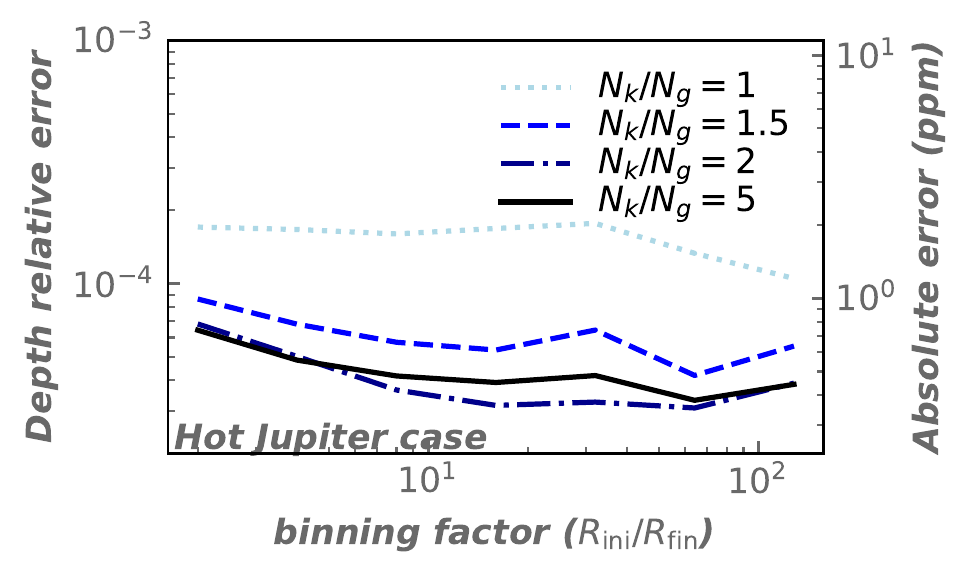} }
\caption{Numerical relative error introduced by the binning process as a function of the binning factor (ratio of the initial resolution, $\Res_\mathrm{ini}$, to the final one, $\Res_\mathrm{fin}$) for four different values of the number of intermediate sampling points $\Nk$ (See \sect{algo}; dotted: $\Nk=\Ng$; dashed: $\Nk=1.5\Ng$; dot-dash:  $\Nk=2\Ng$; solid: $\Nk=5\Ng$, where $\Ng=20$). Top: The RMS relative top-of-atmosphere flux error in our Mars-like case. Bottom: The RMS relative transit depth error in our hot-Jupiter case. The right axis gives the absolute error to be compared to the results of \figs{fig:sampling_error}{fig:res_comp}. With a sufficient sampling, the errors are under the $10^{-3}-10^{-4}$ level depending on the geometry, which is below the error introduced by the \corrk1 method in the first place. }
\label{fig:comparison}
\end{figure}

The results of the comparison are shown in \fig{fig:comparison}. 
For the Mars emission case, we compare the RMS standard deviation of the relative difference between the two spectra in the 1-200$\,\mu$m range. 
For the Hot-jupiter in transmission test case, the comparison metric used is the RMS standard deviation of the relative difference between the two spectra in the 1-2$\,\mu$m window.

All the \kcoeffs1 computed in this work use 20 Gauss-Legendre quadrature points in $\g$-space. Because the number of intermediate points used to resample the \gdist1 in opacity space ($\Nk$; see \sect{algo}) is a free parameter, we tried different values of this parameter. We see that an insufficient intermediate sampling results in significant numerical errors, as could be anticipated. However, when $\Nk$ is a factor of a few larger than the initial number of $g$-points, the accuracy of the method asymptotes below $10^{-3}-10^{-4}$, which is itself much lower than the error introduced by the \corrk1 method in the first place (see for example \citealt{ATM17}). We note that although choosing a larger $\Nk$ may slightly increase the binning time, this does not affect the efficiency of the subsequent radiative transfer calculations in any way. We thus used results with $\Nk=5\Ng$ as a baseline. We also found that using initial data with 40 Gauss-Legendre points did not significantly affect the performances of the binning procedure. Our algorithm is thus robust to the initial choice of $\g$-space sampling.

The spectra obtained with this set of binned-down \kcoeffs1 are also compared directly to the reference high-resolution spectrum as described in \sect{sec:sampling} to give the absolute difference shown in gray with stars in \fig{fig:sampling_error}. Although not strictly equal because they are computed using a slightly different reference spectrum, the absolute difference shown in \fig{fig:sampling_error} and the relative difference from \fig{fig:comparison} can be shown to be consistent by multiplying the latter by the average transit depth as shown on the right axis of \fig{fig:comparison}.
In this particular case, we further wanted to test the impact of non-integer binning. For this a new set of \kcoeffs1 were computed in the exact same way, but with a spectral grid that is slightly shifted in wavenumber to force non-integer binning. As shown by the dotted gray curve in \fig{fig:sampling_error}, the accuracy is lower, although this method remains more accurate than sampled cross sections. In addition, the error decreases with the binning factor because the errors at the boundaries of our spectral bins are more diluted.

We therefore conclude that our binning method does not introduce any significant additional errors. 

\section{Conclusions}\label{sec:conclusion}

We have demonstrated mathematically how the \kdist1 of the opacity in a large band can be obtained from the $k$-distributions inside each of the smaller bands that it encompasses. Then we have presented a numerical algorithm to use this property to bin down precomputed tables of \kcoeffs1 to any arbitrary resolution. Finally, we showed, in two concrete cases, that the numerical error added by this procedure is relatively small compared to the errors due to the more general use of the \corrk1 approach. 

To facilitate the implementation of this new method, we developed \verb"Exo_k", an open-source Python library that has been designed to handle many different sources of radiative data (including, of course, \kcoeff1 tables, but also cross section and collision-induced-absorption tables). Because this library is still rapidly evolving, see our extensive online documentation for tutorials, example notebooks, and a complete description of the library's features\itref{exok_foot}.

We think the flexibility that this approach brings to the \corrk1 method opens up a completely new way of disseminating reference radiative data for use in atmospheric models: precomputed \kcoeffs1 can now be directly distributed at a fraction of the data volume while providing fast and accurate results for any user that does not need a higher resolution. For example, $\Res=1000$ \kcoeffs1 computed on 20 gauss points such as that distributed by the ExoMol project take about 50 times less space on disk than the high-resolution spectra used to produce them. This is a considerable compression factor, and that does not include the hundreds of CPU hours that would be needed to compute the spectra.   

This of course makes us wonder what would be the optimum resolution and quadrature for a flexible and versatile \corrk1 database. Our test show that $\Res=1000$ \kcoeffs1 computed on 20 gauss points, when combined with the flexibility brought by a tool like \verb"Exo_k", seem to be sufficient for most 1D-3D atmospheric model applications\footnote{Very cold and low pressure atmospheres, such as Pluto's, may of course require special care because of their very narrow lines, which usually call for tailored quadrature rules.} as well as modeling synthetic spectroscopic observations of planetary atmospheres up to that resolution. This method can thus be used to model HST, Spitzer, Ariel \citep{TDE18}, and low-resolution JWST data. For high-resolution JWST observation modes, higher resolution radiative data will be needed. 

\begin{acknowledgements}
I thank F. Selsis for being, as always, an outstanding scientific sparring partner and the TauREx team at UCL for pointing out many important Python tricks and libraries. I am also indebted to K. Chubb and M. Turbet for making some radiative data available before their publication. This project has received funding from the European Research Council (ERC) under the European Union's Horizon 2020 research and innovation programme (grant agreement n$^\circ$ 679030/WHIPLASH).
\end{acknowledgements}

\bibliography{biblio}
\bibliographystyle{aa}


\appendix

\end{document}